\documentclass[letter,superscriptaddress,twocolumn,prl,showkeys,showpacs]{revtex4-1}

\usepackage[utf8]{inputenc}
\usepackage{amsmath}   
\usepackage{graphicx}   
\usepackage{verbatim}  
\usepackage{color}     
\usepackage{subfigure}  
\usepackage{hyperref}   
\usepackage{natbib}
\usepackage{gensymb }
\usepackage{color}
\usepackage{epstopdf}
\usepackage{multirow}
\usepackage{xfrac}

\begin{document}

\title{Local resilience of the $1T$-TiSe$_2$ charge density wave to Ti self-doping}

\author{B. Hildebrand}
\altaffiliation{Corresponding author.\\ baptiste.hildebrand@unifr.ch}
\affiliation{D{\'e}partement de Physique and Fribourg Center for Nanomaterials, Universit{\'e} de Fribourg, CH-1700 Fribourg, Switzerland}

\author{T. Jaouen}
\altaffiliation{Corresponding author.\\ thomas.jaouen@unifr.ch}
\affiliation{D{\'e}partement de Physique and Fribourg Center for Nanomaterials, Universit{\'e} de Fribourg, CH-1700 Fribourg, Switzerland}

\author{C. Didiot}
\affiliation{D{\'e}partement de Physique and Fribourg Center for Nanomaterials, Universit{\'e} de Fribourg, CH-1700 Fribourg, Switzerland}

\author{E. Razzoli}
\affiliation{D{\'e}partement de Physique and Fribourg Center for Nanomaterials, Universit{\'e} de Fribourg, CH-1700 Fribourg, Switzerland}

\author{G. Monney}
\affiliation{D{\'e}partement de Physique and Fribourg Center for Nanomaterials, Universit{\'e} de Fribourg, CH-1700 Fribourg, Switzerland}

\author{M.-L. Mottas}
\affiliation{D{\'e}partement de Physique and Fribourg Center for Nanomaterials, Universit{\'e} de Fribourg, CH-1700 Fribourg, Switzerland}

\author{F. Vanini}
\affiliation{D{\'e}partement de Physique and Fribourg Center for Nanomaterials, Universit{\'e} de Fribourg, CH-1700 Fribourg, Switzerland}

\author{C. Barreteau}
\affiliation{Department of Quantum Matter Physics, University of Geneva, 24 Quai Ernest-Ansermet, 1211 Geneva 4, Switzerland}

\author{A. Ubaldini}
\affiliation{Department of Quantum Matter Physics, University of Geneva, 24 Quai Ernest-Ansermet, 1211 Geneva 4, Switzerland}

\author{E. Giannini}
\affiliation{Department of Quantum Matter Physics, University of Geneva, 24 Quai Ernest-Ansermet, 1211 Geneva 4, Switzerland}

\author{H. Berger}
\affiliation{Institut de G{\'e}nie Atomique, Ecole Polytechnique F{\'e}d{\'e}rale de Lausanne, CH-1015 Lausanne, Switzerland}

\author{D. R. Bowler}
\affiliation{London Centre for Nanotechnology and Department of Physics and Astronomy, University College London, London WC1E 6BT, UK}

\author{P. Aebi}
\affiliation{D{\'e}partement de Physique and Fribourg Center for Nanomaterials, Universit{\'e} de Fribourg, CH-1700 Fribourg, Switzerland}

\begin{abstract}

In Ti-intercalated self-doped $1T$-TiSe$_2$ crystals, the charge density wave (CDW) superstructure induces two nonequivalent sites for Ti dopants. Recently, it has been shown that increasing Ti doping dramatically influences the CDW by breaking it into phase-shifted domains. Here, we report scanning tunneling microscopy and spectroscopy experiments that reveal a dopant-site dependence of the CDW gap. Supported by density functional theory, we demonstrate that the loss of the long-range phase coherence introduces an imbalance in the intercalated-Ti site distribution and restrains the CDW gap closure. This local resilient behavior of the 1$T$-TiSe$_2$ CDW reveals a novel mechanism between CDW and defects in mutual influence.     

\end{abstract}
\date{\today}
\maketitle

The quasi-two-dimensional transition-metal dichalcogenide (TMDC) 1$T$-TiSe$_2$ has been largely studied over many years with the desire to understand the mechanisms lying behind its many interesting properties related to its phase transitions. Below $T_{\text{CDW}}\approx 200$ K, 1$T$-TiSe$_2$ exhibits a commensurate CDW phase with a 2x2x2 modulation and a weak periodic lattice distortion (PLD) \cite{salvo1976}. Upon Cu intercalation \cite{Morosan2006a}, and under pressure \cite{Kusmartseva2009a}, it can also host superconductivity that has been proposed to emerge in incommensurate CDW domain walls therefore reflecting the complex 1$T$-TiSe$_2$ phase diagram \cite{Joe2014, Li2015}. 

Doping has shown to be an important tuning parameter of these collective mechanisms \cite{salvo1976, Morosan2006a, Hildebrand2016}. In particular, intercalation of Ti dopants, known to occur depending on the crystal growth temperature \cite{salvo1976}, leads to electron-donor impurity states close to the Fermi energy \cite{Hildebrand2014}, enhances the Coulomb screening, and tends to reduce long-range electronic correlations. In a recent scanning tunneling microscopy (STM) study of Ti self-doped 1$T$-TiSe$_2$ crystals, it has been further reported that for sufficient dopant concentration, the CDW breaks up in randomly phase-shifted domains with subsisting commensurate 2x2 charge modulation separated by atomically sharp phase slips \cite{Hildebrand2016}. This first observation of short-range phase coherent CDW nanodomains induced by Ti-doping not only provides new insight about the microscopic nature of the 1$T$-TiSe$_2$ CDW, but is also of great concern for the understanding of the interplay between dopants and novel electronic phases of TMDCs, in general \cite{Bhoi2016, Liu2016, Ang2015, Joe2014, Ang2012, Morosan2010, Morosan2006a, Fang2005}.

\begin{figure}[b]
\includegraphics[scale=1]{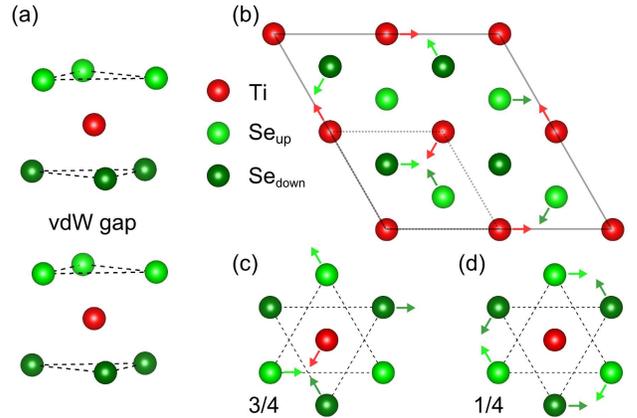}
\caption{\label{fig1} (Color online) (a) Structural model of 1$T$-TiSe$_2$. (b)  Top view (along the $c$-axis) of the PLD accompanying the 1$T$-TiSe$_2$ CDW phase transition (the atomic displacements are represented by arrows). The dotted and full black lines represent the unit cells of the normal phase and of the CDW superstructure, respectively. (c), (d) Top views of the 1$T$-TiSe$_2$ atomic structure showing the PLD displacements with $\sfrac{3}{4}$ of TiSe$_6$ octahedra distorted by Ti-Se bond shortening (c), and $\sfrac{1}{4}$ concerned by a rotation of the Se atoms around a non-moving Ti atom (d).} 
\end{figure}

\begin{figure*}[t]
\includegraphics[scale=1]{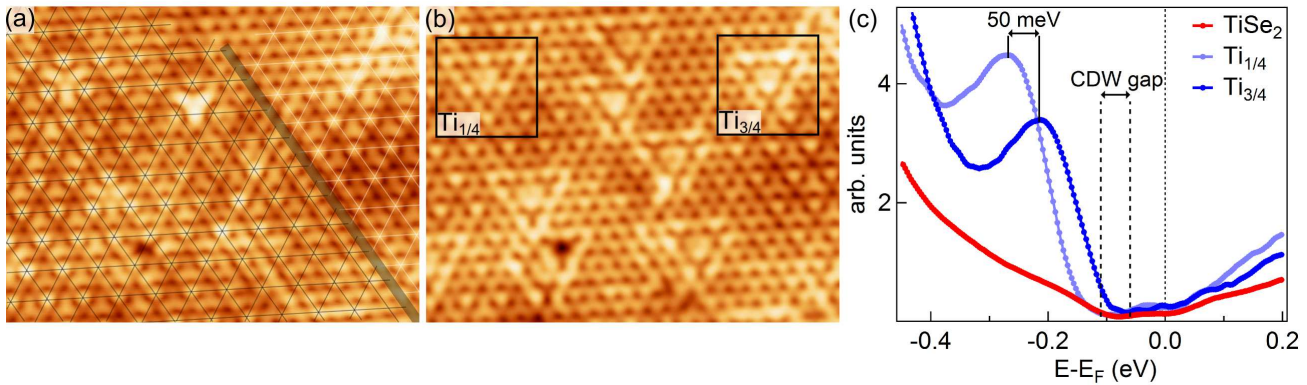}
\caption{\label{fig2} (Color online) (a) 8$\times$6 nm$^2$ constant current STM image of the 860 \celsius$~$-grown 1$T$-TiSe$_2$ crystal showing the CDW charge modulation within two phase-shifted CDW domains separated by one phase-slip. Two meshes corresponding to the two phase-shifted CDW modulations are superimposed and the phase-slip location is indicated by the full grey line. V$_{\text{bias}}$= -50 mV, $I$= 0.2 nA. (b) Constant current STM image at $100$ mV bias voltage and $0.2$ nA current set point of the same region as (a) highlighting the characteristic fingerprints of the $\sfrac{1}{4}$  and $\sfrac{3}{4}$ intercalated-Ti conformations in the presence of CDW (black squares on the left and right-hand sides, respectively). (c) $dI/dV$ curves on top of the two intercalated-Ti conformations (light and dark blue curves, each of them averaged over 5 spectra) and on defect-free region (red curve, averaged over 18 spectra).} 
\end{figure*}

In this paper, we report STM and scanning tunneling spectroscopy (STS) experiments that allow site-specified probing of the local density of states (LDOS) close to the Fermi level (E$_F$). We demonstrate that the loss of the long-range phase coherence is a local resilient behavior of the CDW to self-doping. Together with density functional theory (DFT) calculations, our observations show that the CDW locally adapts to the random distribution of the intercalated-Ti atoms existing in two nonequivalent PLD-related conformations. This CDW-induced Ti-conformation imbalance not only explains the CDW's domain formation in Ti self-doped 1$T$-TiSe$_2$, but also reveals a novel CDW-impurities cooperative mechanism.

The 1$T$-TiSe$_2$ single crystals were grown at 860 \celsius$~$ by iodine vapor transport, therefore containing 1.21$\pm$0.14$\%$ of intercalated-Ti atoms \cite{Hildebrand2016}. The base pressure was better than 5$\times$10$^{-11}$mbar and constant current STM images were recorded at 4.7 K using an Omicron LT-STM, with bias voltage V$_{\text{bias}}$ applied to the sample. The differential conductance $dI/dV$ curves (STS) were recorded with an open feedback loop using the standard lock-in method.

DFT model calculations were performed using the plane-wave pseudopotential code VASP \cite{Kresse1993, Kresse1996}, version 5.3.3. Projector augmented waves \cite{Kresse1999} were used with the Perdew-Burke-Ernzerhof (PBE) \cite{Perdew1996} exchange correlation functional. The cell size of our model was 28.035 \AA~$\times$ 28.035 \AA. The 1$T$-TiSe$_2$ surface was modeled with two layers and the bottom Se layer fixed. A Monkhorst-Pack mesh with 2$\times$2$\times$1 $k$ points was used to sample the Brillouin zone of the cell. The parameters gave an energy difference convergence of better than 0.01 eV. During structural relaxations, a tolerance of 0.03 eV/\AA~ was applied.

In pristine 1$T$-stacked TiSe$_2$ [Fig. \ref{fig1}(a)], there are two possible PLD-related sites for the eight Ti structural atoms of the 2x2x2 CDW unit cell. Indeed, six of them experience a displacement inducing a Ti-Se bond shortening [referred as $\sfrac{3}{4}$ conformation, Fig. \ref{fig1}(c)], whereas the two remaining ones are concerned by a rotation of the six neighboring Se atoms that only slightly changes the Ti-Se interatomic distances [referred as $\sfrac{1}{4}$ conformation, Fig. \ref{fig1}(d)].

In Ti self-doped 1$T$-TiSe$_2$ crystals, Ti defects are intercalated in the van der Waals (vdW) gap directly in line with the structural Ti atoms [Fig. \ref{fig1}(a)] \cite{Hildebrand2014}. Therefore, the CDW phase transition intrinsically implies two nonequivalent sites in the unit cell for each Ti defect, with also three times more $\sfrac{3}{4}$ sites than $\sfrac{1}{4}$ ones \cite{Novello2015}. Since the six TiSe$_6$ octahedra concerned with Ti-Se bond shortening are energetically favorable with respect to the undistorted structure, and the two rotating octahedra are not \cite{whangbo1992}, it becomes highly desirable to investigate the impact of nonequivalent Ti dopants sites on the local electronic properties of the CDW.
  
\begin{figure*}[t]
\includegraphics[scale=1]{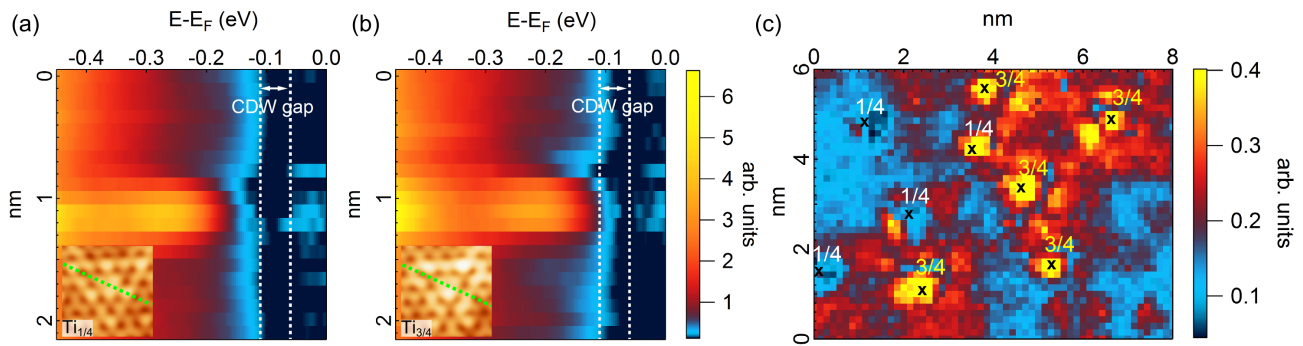}
\caption{\label{fig3} (Color online) (a), (b) Position-dependent intensity plots of the LDOS between -450 mV and $E_F$. Ten $dI/dV$ curves have been measured as a function of position with an interval of 1.4 \AA, through the $\sfrac{1}{4}$ and $\sfrac{3}{4}$ intercalated-Ti defects as indicated by the dotted green lines in the insets. Identical color scales have been used in (a) and (b). The white dashed lines on (a) and (b) mark the CDW gap region between $\sim$-110 and $\sim$-60 mV. (c) Real-space $dI/dV$ map obtained at V$_{\text{bias}}$= -110 mV on the same region as Fig. \ref{fig2}(a) and \ref{fig2}(b) showing the real-space CDW gap perturbation due to intercalated-Ti atoms. The positions of $\sfrac{1}{4}$ and $\sfrac{3}{4}$-conformed Ti defects are also indicated.}
\end{figure*}

Figures \ref{fig2}(a) and \ref{fig2}(b) present 8$\times$6 nm$^2$ filled and empty-state  STM images of the 1$T$-TiSe$_2$ surface respectively recorded at -50 mV and +100 mV. The high resolution filled-state STM image close to E$_F$ [Fig. \ref{fig2}(a)] clearly shows the atomically-resolved CDW superstructure and the presence of one atomically-sharp phase slip [gray line Fig.\ref{fig2}(a)], separating two commensurate CDW domains as expected from the used Ti-doping density and associated CDW domain sizes \cite{Hildebrand2016}. The nonequivalent intercalated-Ti atoms are best resolved and differentiated at positive V$_{\text{bias}}$ \cite{Novello2015}. Empty-state STM images [Fig. \ref{fig2} (b)] reveal their characteristic non-identical patterns consisting of two concentric bright triangles centered on the defect and CDW maxima locations with respect to the underlying Se atomic layer that depend on their $\sfrac{1}{4}$ or $\sfrac{3}{4}$ conformations \cite{Novello2015}.

Figure \ref{fig2}(c) shows tunneling spectra at specified sites of the 1$T$-TiSe$_2$ surface. Each tunneling curve has been measured at different but equivalent positions of the surface and have shown to be highly reproducible \footnote{The STS curves of Figure \ref{fig2}(c) have been renormalized on the basis of the STM topography image made in parallel to the STS map for correcting the exponential decay of the tunneling current associated to the different tip-surface distances on top of a Ti defect.}. In defect-free surface regions, the LDOS corresponds to the one of the CDW state in an overall electron-doped 1$T$-TiSe$_2$ crystal [red curve on Fig. \ref{fig2} (c)]. Indeed, the DOS is non zero at $E_F$ and shows a slight drop of the electron density that corresponds to the Ti 3$d$ conduction band in the CDW state. The CDW gap [indicated by the arrow between the dashed lines Fig. \ref{fig2} (c)], separating the occupied Se 4$p$ orbitals and the Ti 3$d$ electron pocket, opens below $E_F$ between $\sim$-110 and $\sim$-60 mV \cite{Monney2009a}.

The nonequivalent nature of the intercalated-Ti atoms is also clearly manifested in the filled-state part of the STS curves made on top of the $\sfrac{1}{4}$ and $\sfrac{3}{4}$ conformations [light and dark blue curves on Fig. \ref{fig2} (c)]. Indeed, $\sim$200-300 mV below $E_F$, the induced electron-donor impurity states are energetically non-degenerate with the one associated to the $\sfrac{1}{4}$-conformed Ti defects lower in energy by $\sim$50 mV [arrow between the full black lines on Fig. \ref{fig2} (c)] in such a way that the $\sfrac{3}{4}$ impurity state starts to overlap the CDW gap below $E_F$. 

Figures \ref{fig3}(a) and \ref{fig3}(b) show position-dependent intensity plots of the LDOS below $E_F$. The associated $dI/dV$ curves have been measured as a function of position through the $\sfrac{1}{4}$ and $\sfrac{3}{4}$ intercalated-Ti defects, respectively, with an interval of 1.4 \AA~ [dotted green lines in the insets of Fig \ref{fig3} (a), (b)]. They clearly demonstrate the spatial variation of the LDOS and allow to compare the extension of the CDW perturbations introduced by the nonequivalent Ti dopants.

At the intercalated-Ti sites [profiles with highest intensity Fig \ref{fig3} (a), (b)], the LDOS associated with the $\sfrac{1}{4}$ and $\sfrac{3}{4}$ conformations shows the electron-donor impurity states that both develop $\sim$200-300 mV below $E_F$ on a relatively short length scale. Although at these energies they introduce electronic perturbations of limited spatial extension that relates to the localized nondispersing nature of intercalated-Ti 3$d$ orbitals, the impact of their PLD-related conformations on the CDW gap is nonequivalent [CDW gap region between $\sim$-110 and $\sim$-60 mV as indicated by the white dashed lines on Fig. \ref{fig3}(a) and \ref{fig3}(b)]. Indeed, the $\sfrac{3}{4}$ impurity state overlaps the CDW gap [Fig. \ref{fig3}(b)] much more than the $\sfrac{1}{4}$-conformed one [Fig. \ref{fig3}(a)]. The $dI/dV$ map obtained at V$_{\text{bias}}$= -110 mV [Fig. \ref{fig3} (c)], i.e. at the occupied edge of the CDW gap, shows an higher DOS at and around the $\sfrac{3}{4}$ intercalated-Ti atoms therefore indicating a more extended real-space electronic perturbation of the CDW gap as well as of defect states in their immediate vicinity [see $\sfrac{1}{4}$-conformed Ti defect at coordinates (3.6, 4.2) nm on Fig. \ref{fig3} (c)]. This demonstrates that the $\sfrac{3}{4}$-conformed Ti defects have an overall detrimental effect on the CDW state and points towards their higher capability to destroy the CDW order. 

\begin{figure}[t]
\includegraphics[scale=1]{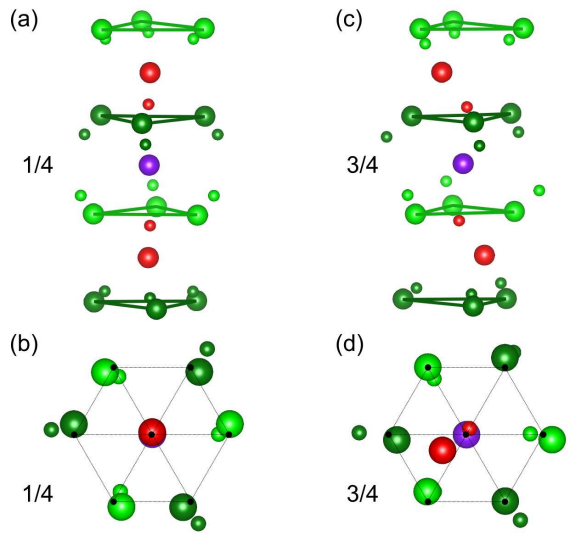}
\caption{\label{fig4} (Color online) DFT-calculated relaxed structures of the 2x2x2 1$T$-TiSe$_2$ lattice with (small spheres) and without (large spheres) intercalated-Ti atoms (violet spheres). Two TiSe$_6$ octahedra separated by one vdW gap are considered. (a)-(b), (c)-(d), Side and top views (top TiSe$_6$ octahedron along the $c$-axis) for the $\sfrac{1}{4}$, respectively $\sfrac{3}{4}$ Ti-defect conformation. The amplitude of displacements induced by Ti intercalation have been enhanced by a factor of 10. The black dots in (b) and (d) indicate the atomic positions in the non-CDW cell, i.e. without PLD.}
\end{figure}

Interestingly, statistics made on the relative density of the $\sfrac{1}{4}$ and $\sfrac{3}{4}$ Ti defect conformations finally provide an additional key point for understanding the occurrence of CDW phase-slips, i.e. the loss of the long-range phase coherence of the CDW state in Ti self-doped 1$T$-TiSe$_2$ crystals. It is known that increasing Ti doping dramatically influences the CDW by breaking it into phase-shifted domains \cite{Hildebrand2016}. Whereas for low Ti-doping, the CDW exhibits a long-range order and a perfect distribution of the $\sfrac{1}{4}$ (25\%) and $\sfrac{3}{4}$ (75\%) Ti-defect sites \cite{Novello2015}, for the studied highly-doped crystal, a careful analysis of 65 intercalated-Ti atoms in the presence of the CDW modulation shows that more than 35$\%$ of them have a $\sfrac{1}{4}$ conformation [instead of 25$\%$ as clearly depicted Fig. \ref{fig2}(b) and Fig. \ref{fig3}(c) where already four out of nine Ti defects are in $\sfrac{1}{4}$ conformation].

This finding indicates that intercalated-Ti defects do not act as a fixed random landscape that can be described by an unique perturbation potential, as it is the case in the well-known impurity pinning theory initially proposed by McMillan \cite{McMillan1975}. Rather, their PLD-induced conformation is chosen by the nucleating CDW and phase-slips allow the CDW lattice to accommodate the random defect distribution by favoring the $\sfrac{1}{4}$ Ti conformation. In other words, the local resilient behavior of the 1$T$-TiSe$_2$ CDW to Ti-doping constitutes a novel mechanism between CDW and defects in mutual influence.   

Finally, we would like to address the crucial role of electron-lattice interaction that lies behind the induced imbalance in intercalated-Ti conformations and that is closely related to the nonequivalent CDW gap perturbation. DFT calculations performed for the two nonequivalent Ti defects in the vdW gap of the PLD-distorted 1$T$-TiSe$_2$ lattice confirm that the total energy is lower for the $\sfrac{1}{4}$ conformation (70 meV per intercalated-Ti atom). Looking at the associated structures after relaxation [Fig. \ref{fig4}], we first see that for the $\sfrac{1}{4}$ conformation [Fig. \ref{fig4} (a)], the lattice distortion is very close to the one calculated for a non-CDW cell \cite{Hildebrand2016}. The atomic displacements are fully symmetric with respect to the $\sfrac{1}{4}$ intercalated-Ti defect [Fig. \ref{fig4} (b)] therefore leaving unchanged the three-fold symmetry of the CDW as confirmed by our STM measurements [inset Fig. \ref{fig3} (a)]. 

On the contrary, as experimentally observed [inset Fig. \ref{fig3} (b)], the lattice distortion induced by $\sfrac{3}{4}$-conformed Ti defects clearly breaks the CDW three-fold symmetry [Fig. \ref{fig4} (c), (d)], with a calculated magnitude of displacements 11$\%$ larger than the one calculated in the $\sfrac{1}{4}$ cell. Indeed, the $\sfrac{3}{4}$-conformed Ti defects induce large lateral displacements of the two neighboring Ti structural atoms [Fig. \ref{fig4} (c), (d)], in such a way that the energetically favorable TiSe$_6$ octahedra concerned with Ti-Se bond shortening (2.49 \AA, as obtained by DFT) in pristine $1T$-TiSe$_2$, undergo a Ti-Se bond elongation of 2 \% (2.54 \AA) under Ti intercalation.  

Based on our STM/STS measurements and DFT calculations, we can thus draw the following conclusions. The atomic displacements as well as the electronic perturbation of the CDW gap induced by the $\sfrac{3}{4}$ intercalated-Ti conformation are larger than those induced by the $\sfrac{1}{4}$ one. The fact that the CDW avoids to perturb the PLD-related $\sfrac{3}{4}$ conformation with a Ti impurity by favoring the $\sfrac{1}{4}$ defect conformation is evidence for the predominant role of the Ti-Se bond shortening for the stabilization of the modulated structure of $1T$-TiSe$_2$, as proposed in pseudo Jahn-Teller scenario \cite{whangbo1992}. Therefore, with the increase of the intercalated-Ti density, it becomes energetically favorable to introduce CDW domain boundaries since the cost in elastic energy starts to be compensated by the electronic energy gain brought by the induced Ti-conformations imbalance. 

To summarize, our site-specified STM/STS study of Ti-doped 1$T$-TiSe$_2$ crystals allows a deep investigation of the impact of nonequivalent dopant sites on the local electronic properties of the CDW. The ability of the CDW state to locally determine the PLD-related conformation of intercalated atoms not only constitutes a novel CDW-impurities cooperative mechanism but also highlights the underlying role of the dopant landscape for the CDW phase patterning resulting in  spatially modulated electronic states that are expected to be a key ingredient for the emergence of mixed CDW-superconducting phases in TMDCs \cite{Li2015}.

\begin{acknowledgments}
This project was supported by the Fonds National Suisse pour la Recherche Scientifique through Div. II. We would like to thank C. Monney, H. Beck, A. M. Novello, and C. Renner for motivating discussions. Skillful technical assistance was provided by F. Bourqui, B. Hediger and O. Raetzo.

B.H. and T.J. equally contributed to this work.
\end{acknowledgments}

\bibliography{library1}
\end{document}